\documentstyle[psfig]{article}
\begin{document}
\begin{sloppypar}
\newcommand{\be}{\begin{equation}}
\newcommand{\ee}{\end{equation}}
\large
\begin{center}

{\bf Dielectric properties of multiband electron systems: 

II - Collective modes}

\vspace{5mm}

 P. \v{Z}upanovi\'{c}
 
{\em Department of Physics,  
Faculty of Science and Art, University of Split,  
Teslina 10, 21000 Split, Croatia} \vspace{5mm}\\
 A. Bjeli\v{s} and  S. Bari\v{s}i\'{c}

{\em Department of Physics,  
Faculty of Science, University of Zagreb,  
P.O.B. 162, 

10001 Zagreb, Croatia}
\end{center}
\vspace{5mm}

\begin{center}
{\bf Abstract}
\end{center}
Starting from the tight-binding dielectric matrix  in the random phase approximation we
examine the collective modes and electron-hole excitations in a two-band electronic system. 
For long wavelengths (${\bf q}\rightarrow0$), for which most of the analysis is carried out, 
the properties of the collective modes  are closely related to the symmetry of the atomic
orbitals involved in the tight-binding states.
In insulators there are only inter-band charge oscillations. If atomic dipolar transitions are
allowed, the corresponding collective modes reduce in the asymptotic limit of vanishing 
bandwidths to Frenkel excitons for an atomic insulator with weak on-site interactions. 
The finite bandwidths renormalize the dispersion of these modes and 
introduce a continuum of incoherent inter-band electron-hole excitations. 
The possible Landau damping of collective modes due to the presence of this continuum 
is discussed in detail. 

In conductors the intra-band charge fluctuations give rise to plasmons. If the atomic dipolar 
transitions are forbidden, the coupling of inter-band collective modes and plasmons tends to 
zero as ${\bf q}\rightarrow0$. On the contrary, in dipolar conductors this coupling is strong and 
nonperturbative, due to the long range monopole-dipole interaction between intra-band and 
inter-band charge fluctuations. The resulting collective modes are hybrids of intra-band 
plasmons and inter-band dipolar oscillations. It is shown that the frequency of the lower 
hybridized longitudinal mode is proportional to the frequency of the transverse dipolar mode
when the latter is small. The dielectric instability in a multi-band conductor is therefore
characterized by the simultaneous softening of a transverse and a longitudinal mode,
which is an important, directly measurable consequence of the present theory.

\bigskip


\bigskip


\mbox{}
\newpage
\section{Introduction}

The present paper completes the analysis of the dielectric screening for a two band system,
derived in the preceding paper I (\cite{zbb1}). Using the tight-binding (TB) scheme, we 
showed that the dielectric matrix $[\varepsilon]$ reduces to a simple $2\times2$ form.
The determinant of this matrix, $\epsilon_m$,  has the significance of the microscopic  
dielectric function. In particular, its zeros and poles in the $\omega$-plane define all collective
and incoherent (electron-hole) excitations, respectively. Here we examine these excitations for two-band
insulators and metals, emphasizing in particular two aspects; the Landau damping of the 
collective modes caused by the inter-band electron-hole continua, and the hybridization 
of the intra-band and inter-band collective modes in the dipolar conductors.  

The RPA dielectric matrix for the TB system with two bands is given by
\begin{equation}
\label{a1}
[ \varepsilon(\mbox{\bf q}, \omega)]=\left[
\begin {array}{cccc}
1-V_{0000}(\mbox{\bf q})\Pi_{00}(\mbox{\bf q}, \omega)                              &-V_{0001}(\mbox{\bf q})[\Pi_{01}(\mbox{\bf q}, \omega)+ \Pi_{10}(\mbox{\bf q}, \omega)]                    
                                                                          \\
 -V_{1000}(\mbox{\bf q})\Pi_{00}(\mbox{\bf q}, \omega)                            &1-V_{0011}(\mbox{\bf q})[\Pi_{01}(\mbox{\bf q}, \omega) + \Pi_{10}(\mbox{\bf q}, \omega) ]                                                                                                                                                                  
\end{array}
\right].
\end{equation}
The matrix elements of the bare Coulomb interaction $V_{l_{1}l_{2}l^{'}_{1}l^{'}_{2}}$
and the bubble polarization diagrams $\Pi_{ij}$ are introduced in paper I [see eqs.(I.8)
and (I.9)]. The two diagonal elements in the matrix (\ref{a1}) come from 
the intraband (00,00) and interband (01,01) polarization processes, while the product of
the off-diagonal elements (00,01) and (10,00) introduces a finite mixing between
them. Obviously the latter vanishes for insulators, since then $\Pi_{00} = 0$. For
conductors, the intraband-interband mixing depends in an essential way on the bare
matrix element $V_{0001}$.  More precisely, the product of off-diagonal elements in 
eq.(\ref{a1}) is finite in the long wavelength limit ${\bf q} \rightarrow 0$ if the long range
part of $V_{0001}$ diverges as $q^{-1}$ \cite{zbb1}. For  Bravais lattices, such
divergence takes place when the matrix element for the dipolar transition between two atomic
TB orbitals, $\mbox{\boldmath $\mu$}_{01}$, is finite, and $V_{0001}$ is dominated by the
monopole-dipole term 
\begin{equation}
\label{d1} 
V_{0001}({\bf q})=\frac{4 \pi i e \mu}{a^{3}} \frac{q_{x}}{q^{2}}. 
\end{equation}
If $\mbox{\boldmath $\mu$}_{01}$ is zero, $V_{0001}({\bf q}\rightarrow 0)$
is the sum of the on-site contribution and the long range monopole-quadrupole 
terms which both behave as $q^{0}$, and thus do not compensate the dependence 
$\Pi_{00}({\bf q}\rightarrow0) \sim q^{2}$ in the off-diagonal matrix element (10,00).

Having in mind the above remarks we divide the analysis of the collective and 
electron-hole excitations in the range (${\bf q}\rightarrow 0, \omega\neq 0$)
into two parts.  First, we shall consider in Sect.2 cases for which the condition
\begin{equation}
\label{a2}
\lim_{{\bf q}\rightarrow 0} V_{0001}^{2}\Pi_{00}[\Pi_{01} + \Pi_{10}] = 0,
\end{equation}
is fulfilled for any direction of ${\bf q}$, so that the microscopic dielectric function 
has a factorized form
\begin{equation}
\label{a3}
\epsilon_m = \epsilon_{intra}\epsilon_{inter},
\end{equation}
where $\epsilon_{intra}$ and $\epsilon_{inter}$ are respectively the diagonal 
elements (00,00) and (01,01) in eq.(\ref{a1}). As we have seen above,
the condition (\ref{a2}) covers two interesting cases, namely those of two band insulators 
and of two-band conductors for which the intra-atomic selection rules forbid the interband 
dipolar transitions. In Sect.3, we examine the conditions under which the collective modes
from the Sect.2 become damped due to the crossing of their dispersion lines with the borders
of the inter-band electron-hole continuum. (Some illustrative examples are considered in 
the Appendix). In Sect.4, we discuss the effects of the finite interband-intraband mixing
(\ref{a2}) on the collective modes. Sect.5 contains concluding remarks.

\section{Pure intraband and interband collective modes}

The factor $\epsilon_{intra}$ in eq.(\ref{a3}) is the usual one-band  RPA  dielectric function 
\cite{nozieres,pines}. The corresponding collective excitations are intraband 
plasmons, the collective excitations of the valence band electron gas. Their 
frequencies are finite if the band is metallic, i. e. partially filled.

On the other hand, the spectrum of collective interband excitations is defined by 
zeros of the interband dielectric function $\epsilon_{inter}$. These excitations 
are dispersive, i.e. undamped, provided that the imaginary part of the interband 
polarization diagram 
\begin{eqnarray}
\label{b1}
\Pi_{01}({\bf q},\omega)+\Pi_{10}({\bf q},\omega)&=&
\frac{2}{N} \sum_{{\bf k}} \left\{
\frac{  n_{0}({\bf k})}
{\omega+E_{0}({\bf k})-E_{1}({\bf k}+{\bf q})
+i\eta} \right.\nonumber\\
     & & \left.-\frac{n_{0}({\bf k}+ {\bf q})    }
{\omega+E_{1}({\bf k})-E_{0}({\bf k}+{\bf q})+i\eta} \right\}
\end{eqnarray}
vanishes. In other words, $\omega({\bf q})$, the zeros of $\epsilon_{inter}({\bf q},\omega)$, 
have to satisfy the condition
\begin{equation} 
\label{b2}
\omega({\bf q}) \pm \left[ E_{0}({\bf k})-E_{1}({\bf k}+{\bf q}) \right]
 \neq 0
\end{equation}
for any {\bf k}, and to be the roots of the equation 
\begin{equation} 
\label{b3}
\frac{4}{N}V_{0011}({\bf q}) 
\sum_{{\bf k}}
n_{0}({\bf k})
 \frac{E_{1}({\bf k}+{\bf q})-E_{0}({\bf k})}
{\omega^{2}-\left[E_{1}({\bf k}+{\bf q})-E_{0}({\bf k}) \right]^{2}}=1.
\end{equation}
If the condition (\ref{b2}) does not hold, the corresponding collective excitation
is damped, i.e. it decays into the quasi continuum of  incoherent
interband electron-hole transitions \cite{pines}.

Let us start from the hypothetical crystal with zero bandwidths for which
$E_{1}({\bf k})=E_{1}$, $E_{0}({\bf k})=E_{0}$. The dispersion of  collective interband modes which 
 follows from eq.(\ref{b3}) is then given by
 \be
\omega^{2}(\mbox{\bf q})=E_{g} \left[E_{g}+2n_{e}V_{0011}({\bf q}) \right],
\label{b4}
\ee
where $E_{g}=E_{1}-E_{0}$ is the difference of the band centers,
and $n_{e}$ is the number of valence band electrons per site. 

For small values of ${\bf q}$ we distinguish two qualitatively different types of dispersion 
emerging from eq.(\ref{b4}). If ${\mbox{\boldmath $\mu$}_{01}}= 0$ the leading 
term in $V_{0011}({\bf q})$ is the intra-atomic interaction  $U_{0011}$, 
while the lowest order possible dependence on ${\bf q}$ is quadratic and comes from the 
quadrupole-quadrupole term in the multiple expansion of the inter-atomic interaction
$W_{0011}({\bf q})$. 
This means that in the limit ${\bf q} \rightarrow 0$ the frequency $\omega({\bf q}=0)$
does not depend on the orientation ${\bf q}/|{\bf q}|$, but has a single value which 
is entirely determined by the intra-atomic parameters, and thus represents a
transition between two atomic orbitals.

If ${\mbox{\boldmath $\mu$}_{01} \neq 0}$ the dipole-dipole term in the 
multipole expansion of  $W_{0011}({\bf q})$ is finite at ${\bf q} = 0$, and varies with the 
orientation of ${\bf q}$ [see eq.(I.32) for the particular case of cubic lattice]. As a consequence,
the frequency $\omega({\bf q}\rightarrow 0)$ varies continuously with the 
change of relative orientation between {\bf q} and the fixed direction of 
$\mbox{\boldmath $\mu$}_{01}$. In other words,  the expression  (\ref{b3})
represents a single dipolar excitation band,  in accordance with the reduced crystal 
symmetry introduced by the choice of TB orbitals [model B of paper I].
The limiting upper and lower edges which correspond to the pure 
longitudinal $({\bf q} \parallel \mbox{\boldmath $\mu$}_{01})$ and 
transverse $({\bf q} \perp \mbox{\boldmath $\mu$}_{01})$ dipolar modes 
are given by 
\begin{eqnarray}
\omega_{L}^{2}({\bf q} \rightarrow 0) \equiv \omega_{L}^{2}=
E_{g}(E_{g}+2n_{e}U_{0011}+\frac{16 \pi n_{e} \mu^{2}}{3a^{3}}), \label{b5} \\
\omega_{T}^{2}({\bf q} \rightarrow 0) \equiv \omega_{T}^{2}= 
E_{g}(E_{g}+2n_{e}U_{0011}-\frac{8 \pi n_{e} \mu^{2}}{3a^{3}}).\label{b6}
\end{eqnarray}  
respectively. As $|{\bf q}|$ increases both $\omega_{L}({\bf q})$ and 
$\omega_{T}({\bf q})$ approach the value of $\sqrt{E_{g}(E_{g} +2n_{e}U_{0011})} $, 
i.e. the width of the dipolar excitation band decreases (Fig.1). 
Let us remind that the result (\ref{b5})  for the longitudinal dipolar mode
is valid provided only that the valence band is full. 

It is worthwhile to note that eqs.(\ref{b5},\ref{b6}), written in terms of the effective
polarization $\mu_{eff}^{2} = \mu^{2}E_{g}/E_{g,eff}$ and the gap 
$ E_{g,eff}^{2} = E_{g}^{2}+ 2nE_{g}U_{0011}$  [see eq.(I.52)],
are just the standard expressions for the Frenkel excitons.
Indeed, after the slight generalization to the band structure
which has a cubic symmetry \cite{zbb1}, these expressions reproduce completely 
the longitudinal and the transverse excitonic collective branches for a cubic crystal
\cite{anderson,knox}. Moreover, they include two distinct types of dipolar 
terms treated separately in Ref.\cite{anderson}, i.e. those representing the simultaneous 
excitation and deexcitation of two atomic electron-hole pairs and the 
simultaneous excitations of two electron-hole pairs at different crystal sites.
It is therefore useful to compare the present derivation of the collective
interband modes via RPA  (which is strictly limited to the weak coupling regime), 
with derivations starting from atomic electron-hole pairs \cite{knox}. 

The collective excitonic state with the wave vector {\bf q} is given here by a linear
superposition in which all interband Bloch electron-hole pairs $a_{1}^{\dagger}(\mbox
{\bf k+q})a_{0}(\mbox{\bf k})$ enter with the same weights and phase factors. When
expressed in terms of local atomic orbitals, this superposition has the form
\begin{equation}
\label{b7}
\sum_{\mbox{\bf R}} e^{i \mbox{\bf qR}}
a_{1}^{\dagger}(\mbox{\bf R})a_{0}(\mbox{\bf R}),
\end{equation}
where $a_{0(1)}^{\dagger}({\bf R})$ creates the electron in the orbital state $0(1)$  
at the crystal site ${\bf R}$. The expression (\ref{b7}) is just the standard description of 
Frenkel exciton as the running wave of the atomic electron-hole pairs \cite{knox}.

Actually, the zero band-width case (\ref{b4}) presents a common asymptotic limit 
of  two incompatible physical regimes. On the one side, one may have  the 
on-site electron-hole repulsion $U_{0101}$ which is much larger than the band-widths
$t_{0,1}$. This binds  the electron and the hole into the pair at a given site. 
The corrections to eqs.(\ref{b5},\ref{b6}) and (\ref{b7}) are then of the order 
of $t_{0,1}^{2}/U_{0101}$ \cite{anderson}. Alternatively, in our weak coupling regime, 
$U_{0101}$ is small with respect to the band-widths, which are in turn assumed to be small 
with respect to the band separation $E_{g}$. The corresponding corrections to 
eqs.(\ref{b5},\ref{b6}) and (\ref{b7}) are now of the order of $t_{0,1}^{2}/E_{g}$. 

From the terminological side, the name of Frenkel exciton is nowadays usually attached
to the regime $t_{0,1}/U_{0101} < 1$ \cite{anderson}. Since we are considering
here primarily the limit $t_{0,1}/E_{g} < 1$ at $U_{0101}$ small, we shall
use for the collective mode  (\ref{b5},\ref{b6}) the term "dipolar excitation" 
all throughout the present paper. We note that the above interpretation 
does not agree with that established in numerous early works (e.g. \cite{hor,izu,miy}) 
on collective excitations in insulators, and adopted by Giaquinta et al \cite{giaq} whose
results for a two-band insulator coincide with our expressions (\ref{b5},\ref{b6}). 
In Ref.\cite{giaq} the longitudinal mode (\ref{b5}) is identified as the renormalized "total" 
plasmon (i.e. $\omega_L^2 = E_g^2 + \frac{2}{3}\omega_{pl,t}^2$),  due to the
simple relation between the frequency of the "total" plasmon and the 
dipolar matrix element, $\omega_{pl,t}^2 = 8\pi \mu^2 n_e E_g/a^{3}$, 
to which the f-sum rule reduces for an insulating two-band system. This
coincidence however does not hold for multiband systems, in which the "total" plasmon 
contains all  interband contributions, while the number of longitudinal collective modes 
increases with the order of the TB dielectric matrix \cite{zbbz}. We emphasize that 
$\omega_{pl,t}$ does not represent the frequency of a collective excitation, 
but appears as a parameter  which characterizes the asymptotic limit $\omega\rightarrow\infty$ 
for $\epsilon_m$. On the other hand, the longitudinal exciton from Refs.[7-10] is 
associated with the exchange and correlation (beyond RPA) contributions
to the screening, and is expected to be inside or below the interband
electron-hole continuum \cite{giaq}. These are precisely the properties of Wannier 
excitons which, as argued in Sect.5,  have to be distinguished from the present dipolar 
collective modes.

\section{Damping of interband collective modes}

In this Section we discuss the interband Landau damping of collective modes, 
following only the more complex case of dipolar excitations (\ref{b5}, \ref{b6}).  
In the limit of zero bandwidths the inter-band electron-hole excitations coincide 
with the inter-atomic transitions between two orbitals, and are represented by 
the horizontal line $\omega=E_{g}$ in Fig.1. For finite bandwidths this line is replaced 
by a region of finite width (\ref{b2}), with the boundaries $E_{m(M)}({\bf q})$ given by
\begin{equation}
\label{c1}
E_{m(M)}({\bf q})=\raisebox{-4mm}{$\stackrel{\displaystyle{\min(\max)}}
{{\bf k}}$}\left[ E_1({\bf k + q}) - E_0({\bf k})\right]
\end{equation}
The illustration of the interband electron-hole 
continuum for particular TB band dispersions is shown in the Appendix.
The collective modes are damped in the region $E_{m}({\bf q}) < \omega < E_{M}({\bf q})$. 
The finite bandwidths which enter into eq.(\ref{c1}) however also alter the dispersion curves 
of the collective modes through eq.(\ref{b3}).  In particular, 
the longitudinal and transverse edges  of the band of collective
modes for a given $|{\bf q}|$ cease to exist when (and if) the renormalised 
energies  $\omega_{L}({\bf q})$ and $\omega_{T}({\bf q})$ cross the lines 
$E_{M}({\bf q})$ and $E_{m}({\bf q})$  respectively. Note that the interband
electron-hole continuum always covers a part of the "interior" of the band
of collective excitations from Fig.1, i.e. there is a finite range of orientations 
of the wave vector  ${\bf q}$ for which the collective dipolar modes are damped.
This is a property of the particular band model B of paper I.

The question which arises now is, whether and under which conditions the 
dispersion curves of  the pure longitudinal and/or transverse collective modes enter 
into the interband electron-hole continuum. The further discussion is
mostly limited to the experimentally interesting range  ${\bf q} \rightarrow 0$.
Note that Figs.1 and 3 suggest that modes from this range are usually 
the last which become damped as the bandwidths increase.

At first, it follows directly from eq.(\ref{b3}) that the crossing of lines 
$\omega_{L}({\bf q})$ and $E_{M}({\bf q})$ is impossible if the sum on the 
left-hand side diverges after replacing $\omega({\bf q})$  by $E_{M}({\bf q})$. The same
is true for the lines $\omega_{T}({\bf q})$ and $E_{m}({\bf q})$, with the replacement 
$\omega({\bf q}) \rightarrow E_{m}({\bf q})$. If this condition holds for all values of 
{\bf q} (with ${\bf q} \parallel \mbox{\boldmath $\mu$}_{01}$ and 
${\bf q} \perp \mbox{\boldmath $\mu$}_{01}$ respectively), the whole line 
$\omega_{L(T)}({\bf q})$ will remain above (below) the  boundary 
$E_{M(m)}({\bf q})$ of the interband  electron-hole continuum. Obviously, 
such argument can be also straightforwardly extended to collective modes 
within the band of Fig.1.
 
The integral in eq. (\ref{b3}) diverges if the locus of zeros of the equation 
\begin{equation} 
\label{c2}
E_{M(m)}({\bf q})-E_{1}({\bf k}+{\bf q})+E_{0}({\bf k})=0 
\end{equation}
in the $d$-dimensional {\bf k}-space has the dimension $d-1$. For 
lower dimensions of the locus the {\bf k}-integration is regular, so that the 
crossing may occur for some finite value of $|{\bf q}|$. As the width of the
interband electron-hole continuum increases the crossing point usually moves 
towards smaller values of $|{\bf q}|$, and eventually reaches the point ${\bf q}=0$.

The properties of  electron-hole locus in the {\bf k}-space are directly linked to the
details in the band dispersions $E_{0(1)}({\bf k})$, in particular to their dimension. 
E.g., for strictly one-dimensional bands the crossing between the dipolar 
collective modes and  the electron-hole lines  is in principle not possible for any 
value of the wave vector. Analogously, the plasmon  dispersion curve $\omega_{pl}({\bf q})$
does not enter into the intraband electron-hole continuum \cite{bloch,zbb2}. For crystals 
with two- and three-dimensional electron bands the locus from eq.(\ref{c2}) is usually a point, 
so that the crossing between the collective modes and the electron-hole lines is not forbidden. 

\section{Hybridization of intraband and interband collective modes}

In the two-band conductors with finite interband dipolar transitions the 
condition (\ref{a2}) is not fulfilled, and the off-diagonal intraband-interband elements 
of the matrix (\ref{a1}) enter into the equation $\epsilon_m = 0$. In order to determine 
analytically collective modes which follow from this equation,  let us consider the limit  
${\bf q}\rightarrow 0$ and assume, like in Sec.2, that the widths of the valence and 
conducting band  are small in comparison with $E_{g}$ and $|E_{g}^{2}- \Omega^{2}|^{1/2}$,  
where $\Omega$ stands for the frequencies of collective modes, yet to be determined. 
We also use the effective mass form of the valence ($l=0$) band dispersion. The intraband
and interband polarization diagrams are then given by
eqs.(I.33) and (I.34) respectively, 
and the equation $ \mbox{Re}\;\epsilon_m({\bf q},\omega)=0$ reduces to 
\begin{eqnarray}
&&\Omega^{4}-\left[E_{g}^{2}+\frac{n_{e}q^{2}}{m^{*}}
V_{0000}({\bf q})+2n_{e}E_{g}V_{0011}({\bf q})\right]\Omega^{2}+\nonumber\\
&& +E_{g}\left[V_{0000}({\bf q})V_{0011}({\bf q})+V_{0001}({\bf q})^{2}
 \right]
\frac{2n_{e}^{2}q^{2}}{m^{*}}
+\frac{n_{e}E_{g}^{2}}{m^{*}}V_{0000}({\bf q})q^{2}=0.
\label{d4}
\end{eqnarray}
where $m^*$ is the effective band mass and $n_e$ is, as before, the number of valence
band electrons per site. Here  it is taken  into account that,  in accordance with the 
previous assumption $t_{i} \ll E_{g}$, the regions of intraband and interband
electron-hole continua do not overlap, i.e. that
\begin{equation}
\mbox{Im} [\Pi_{00}({\bf q},\omega)] \cdot \mbox{Im}
[\Pi_{01}({\bf q},\omega)+\Pi_{10}({\bf q},\omega)]=0.
\label{d5}
\end{equation}

For wave vectors perpendicular to $\mbox{\boldmath $\mu$}$ ($q_{x}=0$) the matrix
element $V_{0001}$ vanishes, so that the longitudinal intraband plasmons are not coupled
to the transverse dipolar modes (\ref{b6}). 
However, the dipolar modes which propagate in other directions,
do couple with plasmons.  This coupling is strongest in the longitudinal direction 
$q=q_{x} \rightarrow 0, q_{y}=q_{z}=0$ which we consider further on. It is
important to note that in this limit the square of the monopole-dipole interaction
(\ref{d1}) and the product  $V_{0000}({\bf q})V_{0011}({\bf q})$ [see eqs. (I.28), (I.31)
and (I.32)] have the same structure and the numerical factors of the same order, so that 
the monopole-dipole interaction cannot be treated perturbatively.

After inserting the bare Coulomb matrix elements, the solutions of eq.(\ref{d4}) read
\begin{eqnarray}
\Omega^{2}_{\pm}=\frac{1}{2}\left\{ \omega_{L}^{2}+\omega_{pl}^{2}  
 \pm  \left[ \left(\omega_{pl}^{2}+\omega_{L}^{2} \right)^{2}
 -4\omega_T^2\omega_{pl}^{2} \right]^{1/2}
\right\}
\label{d7}
\end{eqnarray}
Here $\omega_{pl}=\sqrt{4\pi n_{e}e^{2}/(m^{*}a^{3})}$ is the frequency of the bare 
intraband plasmon. Let us discuss the result (\ref{d7}) by distinguishing between the two opposite
limits, $\omega_{pl}^{2} \gg \omega_{L}^{2}$ and $\omega_{pl}^{2} \ll \omega_{L}^{2}$.
For $\omega_{pl}^{2} \gg \omega_{L}^{2}$ the renormalized frequencies reduce to 
\begin{equation}
\Omega_{+}^{2}=\tilde{\omega}_{pl}^{2} \cong 
\omega_{pl}^{2}+(\omega_L^2 - \omega_T^2)
\left[1 + \frac{ \omega_{T}^{2}}{\omega_{pl}^{2}} \right]
=\omega_{pl}^{2}+\frac{4 \pi n_{e} \alpha}{a^{3}}
\left[1 + \frac{ \omega_{T}^{2}}{\omega_{pl}^{2}} \right] E_{g}^{2}
 \label{d8}
\end{equation}
and
\begin{equation}
\Omega_{-}^{2}= \tilde{\omega}_{L}^{2} \cong
\omega_{T}^{2} \left[1-\frac{\omega_L^2 - \omega_T^2}
{\omega_{pl}^{2}} \right],
\label{d9}
\end{equation}
where  $\alpha=2\mu^{2}/E_{g}$ is the molecular polarizability. It follows from 
eq.(\ref{d8}) that the monopole-dipole interaction increases the effective 
monopole-monopole interaction, causing  an increase of the plasmon frequency 
$\tilde{\omega}_{pl}$. The factor 
$4 \pi n_{e} \alpha /a^{3}[1 +  \omega_{T}^{2} / \omega_{pl}^{2} ]$
is usually of the order of unity in metals. The renormalized plasmon 
frequency  (\ref{d8}) can thus be identified as the so-called 
interband plasmon frequency \cite{chakraverty,barisic2}, usually taken as
equal to $\sqrt{\omega^{2}_{pl}+E_{g}^{2}}$. On the other hand,
the renormalized longitudinal dipolar mode (\ref{d9}) is shifted below
the transverse one. Furthermore, the band of dipolar modes
becomes extremely narrow, i.e. $(\omega_{T}^{2}-\tilde{\omega}_{L}^{2})/
(\omega_L^2 - \omega_T^2) \approx \omega_{T}^{2} / \omega_{pl}^{2} \ll 1$.
Since $\omega_{T} \leq \min (E_{m},E_{M})$, this band, if not damped, 
lies below the interband electron-hole continuum.
  
In the opposite limit $\omega_{L}^{2} \gg \omega_{pl}^{2}$ the frequencies 
 $\Omega_{\pm}$ reduce to
\begin{equation}
\Omega_{+}^{2}=\tilde{ \omega}_{L}^{2} \cong 
\omega_{L}^{2}+\frac{\omega_{pl}^{2}}{\omega_{L}^{2}} 
\left[1+ \frac{\omega_{T}^{2} \omega_{pl}^{2}}{\omega_{L}^{4}} \right]
(\omega_L^2 - \omega_T^2)
\label{d10}
\end{equation}
and
\begin{equation}
\Omega_{-}^{2}=\tilde{ \omega}_{pl}^{2}=
\omega_{T}^{2} \frac{\omega_{pl}^{2}}{\omega_{L}^{2}}\left[1- 
\frac{\omega_{pl}^{2}(\omega_L^2 - \omega_T^2)}{\omega_{L}^{4}}\right].
\label{d11}
\end{equation} 
Within this limit one may distinguish between the two opposite cases, related 
to the ratio $E_{g}^2/(\omega_L^2 - \omega_T^2)$. For 
$E_{g}^2/(\omega_L^2 - \omega_T^2)\gg1$ the renormalized plasmon 
frequency (\ref{d11})  is close to its bare value $ \omega_{pl}$, 
due to $\omega_{L} / \omega_{T} \cong 1$. The opposite limit
represents the regime in which $\omega_{T}$ may be critically softened,
or even  become unstable \cite{chakraverty}. As is seen from eq.(\ref{d11}), 
$\tilde{\omega}_{pl}^{2}$ is proportional to, and smaller than 
$\omega_{T}^{2}$. Note that in both cases the shift of the pure 
longitudinal dipolar mode, given by eq.(\ref{d10}), is small and positive
($\tilde{\omega}_{L}^{2} > \omega_{L}^{2}$).
  
The most important common outcome of the above discussion is the proportionality 
of the frequencies of the lower longitudinal mode $(\Omega_{-})$ and the 
transverse dipolar mode $(\omega_{T})$. Moreover, it follows from eq.(\ref{d7}) 
that in the regime of critical softening the instabilities 
$\omega_{T}\rightarrow0$ and $\Omega_{-}\rightarrow0$ proceed simultaneously,
irrespectively of the relation between $\omega_{pl}$ and $\omega_{L}$, with the ratio
\begin{eqnarray}
(\omega_T/\Omega_-)^2 \simeq 1 + (\omega_L/\omega_{pl})^2 > 1. 
\label{d12a}
\end{eqnarray} 
Thus, after taking properly into account the coupling of the longitudinal dipolar mode and
intraband plasmon, one obtains a common critical behavior for the lower longitudinal mode 
and the transverse mode, although the latter is not coupled to the plasmon. This, to some 
extent unexpected, result brings a new insight into the problem of the instability of dipolar 
collective modes. We note that it is not restricted to the present two band structure \cite{zbbz}.
  
The finite monopole-dipole interaction  also modifies the conclusions of Sect.3,  
concerning the damping of the renormalized collective longitudinal  modes due to 
the electron-hole transitions. Let us keep the assumption that the intraband and 
interband electron-hole continua do not overlap, as expressed by eq.(\ref{d5}).
The  imaginary part of the microscopic dielectric function $\epsilon_{m}$ reads 
\begin{eqnarray}
\mbox{Im} \;\epsilon_m = -\mbox{Im} \;\Pi_{00}V_{0000} [1-V_{0011} \mbox{Re} (\Pi_{01}+\Pi_{10})]
 - \mbox{Im}(\Pi_{01}+\Pi_{10})V_{0011} \times \nonumber\\
 \times (1-V_{0000}  \mbox{Re} \;\Pi_{00})+ 
 V_{0001}^{2}[\mbox{Re}\; \Pi_{00} \mbox{Im} (\Pi_{01}+\Pi_{10})+ 
\mbox{Im} \; \Pi_{00} \mbox{Re} (\Pi_{01}+\Pi_{10})]. 
\label{d13}
\end{eqnarray}
Obviously, without the monopole-dipole interaction the 
intraband plasmons are not damped inside the
$({\bf q},\omega)$ region covered by the interband  electron-hole continuum.
The analogous conclusion holds for the band of dipolar modes, even if it has a
finite overlap with the intraband valence electron-hole continuum.
  
The inclusion of $V_{0001}({\bf q})$ does not change the 
boundaries of the electron-hole continua. However, it brings into 
eq.(\ref{d13}) a new term,  which is finite at all crossings of the 
electron-hole continua and the dispersion curves of longitudinal 
collective modes. Consequently, the longitudinal collective modes become 
damped by entering into both intraband and interband electron-hole continua.	
In particular, this means that the propagation of the renormalized long wavelength 
plasmon [eqs.(\ref{d8}) or (\ref{d11})] is damped within the interband electron-hole 
continua. On the other hand, collective modes with 
finite frequencies at ${\bf q} \rightarrow 0$ are not damped by 
the intraband electron-hole continuum, since the latter 
has an upper boundary, which starts at the origin of the $({\bf q},\omega)$ space.
For finite values of $|{\bf q}|$ this argument does not hold, and 
one has to establish criteria for the absence of damping of longitudinal 
collective excitations along the lines discussed in Sect.3. Note that the pure transverse 
modes exist within the intraband electron-hole continuum,  
since they do not couple to this continuum.

The frequencies (\ref{d7}) are the zeros of the macroscopic dielectric function
[see eq.(I.48)] which is in the absence of the electron-hole damping given by
\begin{equation}
\epsilon_{M}(q_{\parallel},\omega)= \frac{(\omega^{2}-\Omega_{-}^{2})
(\omega^{2}-\Omega_{+}^{2})}
{\omega^{2}(\omega^{2}-\omega_{T}^{2})}.
\label{d14}
\end{equation}
This result includes the well-known special cases of the one band conductor and 
the two-band  insulator. In the former case $\Omega_{-}=\omega_{pl}$ and
$\Omega_{+} \approx \omega_{T}\rightarrow \infty$, so that 
$\epsilon_{M}$ reduces to the RPA expression for the plasmon edge.
In the latter case one recovers, after $ \omega_{pl}=\Omega_{-}=0$ and 
 $\Omega_{+}=\omega_{L}$, the Lorentz-Lorenz expression 
$\epsilon_{M}(q_{\parallel},\omega)=(\omega^{2}-\omega_{L}^{2})/
(\omega^{2}-\omega_{T}^{2})$, i.e. the Lyddane-Sachs-Teller relation
$\epsilon_{M}(\omega=0)/\epsilon_{M}(\omega=\infty)=\omega_{L}^{2}/\omega_{T}^{2}$.
 
Eq.(\ref{d14}) and Fig.2 also clearly show how the simultaneous
instability of the transverse dipolar mode and the lower longitudinal mode in
a conductor is manifested in the macroscopic dielectric properties. This instability 
is preceded by the shifts of both, the zero $(\omega=\Omega_{-})$ and the  pole 
$(\omega=\omega_{T})$ of $\epsilon_{M}(q_{\parallel},\omega)$, 
towards the origin. Note that for the insulator, the zero is at the origin  all the time 
[$\omega=\Omega_{-}=0$], compensating  the metallic $\omega^{-2}$ divergence 
in the expression (\ref{d14}). The instability of the transverse dipolar mode is then 
manifested solely as a shift of the pole at $\omega=\omega_{T}$ towards  the origin.

Eq.(\ref{d14}) and Fig.2 apply also, with a somewhat modified content, to the 
isotropic crystal with three degenerate $p$-bands [see Appendix B in paper I].
The expression (\ref{d14}) is then valid for any orientation of {\bf q}, with the 
frequencies $\omega_{T}$ and $\Omega_{\pm}$ being the true branches of 
collective modes, in contrast to the above anisotropic case in which they represent the edges 
of a continuous band as it is specified in Sect.2.

\section{Conclusion}

The present analysis, although based on the simplest nontrivial example of a multiband system,
leads to some results of  broad significance. First, one may distinguish between
purely intra-band and inter-band collective modes only in the 
long wavelength limit. Even then, this is justified only in the particular cases of an insulator 
and a conductor with forbidden dipolar transitions. Furthermore,  in the zero bandwidth 
limit of the former case the dispersion and the coherent electron-hole distribution of dipolar 
interband modes are just those of Frenkel excitons. We stress again that our identification is at 
variance with that from  earlier literature [7-10], and that  Frenkel 
excitons are not associated only with strong coupling systems \cite{anderson}, but 
are also an asymptotic limit within the RPA approach appropriate for the weak coupling
regime,  provided that the conditions specified in Sect.2 are fulfilled. 

In the case of dipolar conductors there is a finite "off-diagonal" coupling between the intra-band
and inter-band polarization processes. It  is explicitly shown that in the long wavelength
limit this coupling originates from the diverging long-ranged monopole-dipole interaction
which cannot be treated perturbatively. As a consequence, the collective modes are hybridized
from the intra-band (plasmon) and inter-band (dipolar) oscillations. The most interesting 
property of this hybridization is the proportionality of the frequencies of the lower longitudinal 
mode and the (non-hybridized) transverse inter-band mode when both are small.
In other words, the corresponding dielectric instability in conductors proceeds by the 
simultaneous softening of two collective modes. The search for systems in which
phenomenon of this kind takes place would be very desirable. 

It should be pointed out that these results, as well as those regarding the Landau damping of 
the collective modes due to finite regions of incoherent electron-hole continua in the 
$({\bf k}, \omega)$ plane, follow from the TB  microscopic dielectric function (\ref{a1})
which includes properly the most relevant ingredients of the interactions and symmetries 
of a multiband system, and remains transparent to an analytical approach. They could 
not be obtained even approximately from the dielectric function with the intra-band 
and inter-band polarizabilities entering additively, which is the usual form given
in textbooks \cite{pines}. 

The present work deals with the simple (i.e. electron-hole) version of  the RPA,
formulated in paper I.  It can be however extended to the generalized  RPA which
also includes exchange (ladder) contributions \cite{sham,rogan}
to the polarization diagram. Postponing the full account of this calculation 
within the present TB approach for a separate paper, let us mention here that
the spectrum of excitations then acquires  a qualitatively new feature, namely the bottom of 
the interband electron-hole continuum (or continua for more than two bands) 
exhibits a discrete structure. The localised levels, defined by poles of the imaginary part of 
the dielectric function, represent the bound states of interband electron-hole excitations, 
i.e. the Wannier excitons. On the other hand, the dipolar modes from Sects.2 and 4 are 
coherent superpositions of electron-hole pairs, specified by eq.(\ref{b7}).
Hence those excitations and Wannier excitons are two
qualitatively different  types of interband excitations. The former are collective bosonic
dipolar fluctuations, while the latter are the electron-hole states which are localized
in space and may propagate with a finite velocity. There is no a priori reason,
at least in the weak coupling regime, against the coexistence \cite{rogan} of those
excitations, as far as they do not overlap in the $({\bf q},\omega)$ plane.
In the narrow region of overlap the calculation of $[\varepsilon_{m}({\bf q},\omega)]$
would have to be extended to diagrams beyond the exchange RPA. In that respect
the discussion in Sect.3 may be applied only to the continuous part above the discrete
Wannier structure in the region of interband electron-hole excitations.
Finally, additional complications occur in low dimensional systems.

\appendix
\section{ Appendix }

\setcounter{equation}{0}
\renewcommand{\theequation}{A.\arabic{equation}}
\renewcommand{\thesubsection}{A.\Roman{subsection}}

In order to illustrate the discussion of Sect.3 on the crossing between the interband 
collective modes and the boundaries of interband electron-hole continuum, we
consider here two cubic TB bands with bandwidths $t_0$ and $t_1$ and assume 
that the lower (valence) band is either almost empty (A.I) or full (A.II).

\subsection{Nearly empty valence band}

In this case the band dispersions can be approximated by the effective mass form 
$E_{l}(k)=E_{l}-(6-k^{2}a^{2})t_{l}$  with $l=0,1$ and  $k= \mid \mbox{\bf k} \mid $. 
The corresponding electron-hole continuum is sketched in Fig.3. 
The expression (\ref{b3}) then reduces in the limit $\mbox{\bf q} \rightarrow 0$ to 
\begin{equation}
\label{aa2}
\sqrt{E_{m}-\omega}  \arctan\frac{\sqrt{\cal E}}{\sqrt{E_{m}-\omega}}+
\sqrt{E_{m}+\omega} \arctan \frac{\sqrt{\cal E}}{\sqrt{E_{m}+\omega}}=
2\sqrt{\cal E} +\frac{{\cal E}^{3/2}}{3 n_{e} V_{0011}}, 
\end{equation}
with $V_{0011} \equiv V_{0011}({\bf q} \rightarrow 0)$, $E_{m} \equiv E_{g}-
6(t_{1}-t_{0})$, ${\cal E} \equiv (t_{1}-t_{0}) k_{F}^{2}a^{2}$ and  
$k_{F}=(3 \pi^{2} n_{e}N/L^{3})^{1/3}$. Eq.(\ref{aa2}) implicitly determines  the
renormalization of the dipolar collective modes due to finite widths of electronic bands. 
In Fig.4 we show the dependence of the limiting frequencies $\omega_{T}$ and 
$\omega_{L}$ and of the boundaries of the electron-hole continuum 
$E_{m}$ and $E_{M}$ on the difference of band-widths
$\cal E$. Evidently, for any finite $\cal E$ 
a part of the excitation spectrum will be forbidden. In order to organize 
the discussion, it is instructive to distinguish four ranges.

{\em(i)} ${\cal E}>0,\; 0< \omega < {E_{m}}$. 
The left-hand side of eq.(\ref{aa2}) is now bounded between 0 and
 $2\sqrt{\cal E}$. 
The pure transverse dipolar modes exist for 
 ${\cal E}<{\cal E}_{crT}$ where  ${\cal E}_{crT}$ 
is the solution of the equation 
\begin{equation}
\sqrt{2E_{m}}  \arctan\frac{\sqrt{{\cal E}_{crT}}}{\sqrt{2E_{m}}}=
2\sqrt{{\cal E}_{crT}}+\frac{{\cal E}^{3/2}_{crT}}
{3  n_{e}[U_{0011}-4\pi \mu^{2}/(3a^{3})]}. 
\label{aa3}
\end{equation}
At ${\cal E} = {\cal E}_{crT}$ the frequency of the pure transverse mode  becomes equal to the 
lower boundary of the electron-hole  continuum. Putting ${\cal E}_{crT} \ll E_{m}$
in  eq.(\ref{aa3}), in agreement with the assumption that the valence band
is nearly empty, one gets 
$ {\cal E}_{crT}\cong 4 \pi  n_{e}\mu^{2}/a^{3}-3n_{e}U_{0011}$.
 ${\cal E}_{crT}$ is well defined as far as 		
${\cal E}_{crT}/(k_F^2a^2)< [E_{g}+2n_{e}U_{0011}
 -8n_{e} \pi \mu^{2}/(3a^{3})]/6$, 
since only then the pure transverse mode disappears at a finite frequency. 
Otherwise one would have an  instability of the pure transverse mode
$(\omega_{T} \rightarrow 0)$ at the value of   ${\cal E}_{crT}/(k_F^2a^2)$ equal to the 
right-hand side of this inequality.  In considering the Frenkel excitons 
\cite{anderson,knox,davydov} it is usually assumed that the above condition is 
fulfilled. The excitonic instability  has been however invoked recently in the context 
of excitonic mechanism for the superconductivity in high $T_{c}$ systems \cite{chakraverty}.

{\em (ii)} ${\cal E}>0, \;  \omega > E_{M}=E_{m}+{\cal E}$. 
The left-hand side of eq.(\ref{aa2}) is now given by
$\sqrt{\omega-E_{m}} \: \mbox{arth}(\sqrt{{\cal E}}/\sqrt{\omega-E_{m}})+
\sqrt{E_{m}+\omega} \arctan (\sqrt{{\cal E}}/\sqrt{E_{m}+\omega})$.
This function tends to $2\sqrt{{\cal E}}$ for $\omega \rightarrow \infty $, and to 
infinity for $\omega$ approaching the upper boundary of electron-hole 
continuum $E_{M}$. The pure longitudinal 
mode remains outside the electron-hole continuum, and therefore 
does not disappear in the whole region ${\cal E}>0$.
This result is the consequence of the 
fact that the Fermi surface coincides with the surface of the poles 
in the integral of eq.(\ref{b3}). 

{\em (iii)} ${\cal E}<0,\; 0< \omega < E_{M}$. Now $E_{m}$  and $E_{M}$
are the upper and lower  boundaries of the electron-hole continuum respectively.
The function at the left-hand side of eq.(\ref{aa2}) is given by
$\sqrt{E_{m}-\omega} \: \mbox{arth}(\sqrt{\mid {\cal E} \mid}/
\sqrt{E_{m}-\omega})
+\sqrt{E_{m}+\omega} \: \mbox{arth} 
(\sqrt{ \mid{\cal E} \mid}/\sqrt{E_{m}+\omega})$.
The arguments analogous to those in {\em (ii)} lead to the conclusion 
that the pure transverse mode exists in the whole range 
${\cal E}<0$, approaching asymptotically the lower boundary of the 
 electron-hole continuum, as shown in Fig.4.

{\em (iv)} ${\cal E} < 0, \;  \omega > E_{m}$.
The function at the left-hand side of eq.(\ref{aa2}) is now given by
$\sqrt{\omega-E_{m}}  \arctan (\sqrt{\mid {\cal E} \mid}/
\sqrt{\omega-E_{m}})+\sqrt{E_{m}+\omega} \, \mbox{arth} 
(\sqrt{\mid {\cal E} \mid}/\sqrt{E_{m}+\omega})$.
The frequency of the pure longitudinal mode $\omega_{L}$  
touches the upper boundary of 
electron-hole continuum at the frequency $\omega=E_{m}$, i.e. for 
${\cal E}_{crL}$ determined as the zero of the equation 
\begin{equation}
\label{aa5}
\sqrt{2E_{m}} \: \mbox{arth}\frac{\sqrt{\mid {\cal E}_{crL} \mid}}{\sqrt{2E_{m}}}=
2\sqrt{ \mid {\cal E}_{crL} \mid}  -\frac{\mid {\cal E}_{crL} \mid^{3/2}}
{3 n_{e}[U_{0011}+8 \pi \mu^{2}/(3a^{3})]}.
\end{equation}
Again, taking into account that $E_{m} \gg |{\cal E}_{crL}|$, one gets 
$|{\cal E}_{crL}| \cong 8 \pi n_{e} \mu^{2}/a^{3}+3 n_{e}U_{0011}$.

Note  that $\mid{\cal E}_{crL} \mid \cong 2{\cal E}_{crT}+9n_{e}U_{0011}$, 
i.e. on the absolute scale of the bandwidth broadening, the transverse part of the 
band of collective modes disappears before the longitudinal one,  provided that 
$U_{0011} > -{\cal E}_{crT}/(9n_{e})$.

\subsection{Full  (and nearly full) valence band}

When the lower band is full eq.(\ref{b3}) for ${\bf q}=0$ reduces to
\[\frac{4}{N}\left (\frac{L}{2 \pi}\right)^{3}V_{0011}({\bf q}) E_{g}
 \int_{-\pi/a}^{\pi/a} \prod_{i=1}^{3}dk_{i} \times \]
\mbox{}
\begin{equation}
\times \frac{\omega^{2}-E_{g}^{2}+4 {\cal E}^{2}(\sum_{i=1}^{3} \cos k_{i}a)^{2}}
{[\omega^{2}-E_{g}^{2}-4{\cal E}^{2}(\sum_{i=1}^{3} \cos k_{i}a)^{2}]^{2}
 -16 E_{g}^{2} {\cal E}^{2}(\sum_{i=1}^{3} \cos k_{i}a)^{2}}=1. 
\label{aa6}
\end{equation}
The energies of the interband electron-hole transitions with ${\bf q}=0$ lie between 
$E_{m}\equiv E_{g}-6{\cal E}$ and $E_{M}=E_{g}+6{\cal E}$,  
where now ${\cal E}=t_{1} - t_{0}$. From eq.(\ref{aa6}) it follows that for  
$E_{g} \gg 6|{\cal E}|$ the pure transverse mode is damped  for 
$|{\cal E}| > |{\cal E}_{crT}|=3I[4\pi \mu^{2}/(3a^3)-U_{0011}]/(8\pi^{3})$, where 
\begin{equation}
\label{aa7}
 I= \int_{- \pi}^{\pi} \prod_{i=1}^{3} dx_{i} 
\frac{1}{9-(\sum_{i=1}^{3} \cos x_{i})^{2}} .
\end{equation}
The numerical calculation gives $I=41.784$. Due to the reduction of the Fermi
surface to the corners of the first Brillouin zone the pure transverse mode can disappear 
for both ${\cal E}  >0$ and ${\cal E} <0$. Note also that this mode becomes unstable 
$(\omega_{T}=0)$ for  
$|{\cal E}| =E_{g} \sqrt{E_{g}/[8 \pi \mu^{2}/(3a^{3})-2U_{0011}]-2}$. 
 
The analogous analysis performed for pure longitudinal mode shows 
that it becomes damped for both signs of the difference ${\cal E} $ provided that
$|{\cal E}| >|{\cal E}_{crL}|=3I[8\pi \mu^{2}/(3a^3)+U_{0011}]/(8\pi^{3})$.
Note that $|{\cal E}_{crL}| \cong  2 |{\cal E}_{crT}|
 + 9I/(8\pi^3)U_{0011}$, i.e. the pure 
transverse mode disappears before the pure longitudinal one as the difference
increases, provided that $U_{0011} > -{\cal E}_{crT}8\pi^3/(9I)$.
The frequencies of the pure longitudinal and transverse modes 
as well as the lines $E_{m}$ and $E_{M}$ are depicted in the Fig.5.

For the nearly full valence band 
$[k_{F}= \sqrt{3} \;\pi/a-k_{h_{F}}, k_{h_{F}} \ll   \sqrt{3} \;\pi/a]$
the results for ${\cal E}_{crT}$ and  ${\cal E}_{crL}$ 
are modified in a following way. If ${\cal E} >0$
(here again ${\cal E}=t_{1}-t_{0}$), the pure transverse 
mode with ${\bf q} =0$ reaches the value $E_{m}$ for 
${\cal E}_{crT}=3I[4\pi\mu^2/(3a^3) -U_{0011}][1-2 \pi^{3}n_{h}/(9I)]/(8\pi^3)$,
where $ n_{h}=k_{h_{F}}^{3}L^{3}/(3 \pi^{2}N)$. Due to the finite 
Fermi surface which is the locus of electron-hole poles in eq (\ref{b3}),
the frequency of the longitudinal mode remains above the value $E_{M}({\bf q}=0)$. 
 
For ${\cal E} <0$ one has the opposite situation in which the pure 
longitudinal mode disappears for 
$|{\cal E}_{crL}|=3I[8\pi\mu^2/(3a^3)+U_{0011}][1-2 \pi^{3}n_{h}/(9I)]/(8\pi^3)$,
while the pure transverse mode remains below  the electron-hole  continuum.

\newpage
\mbox{}
\vspace{10mm}
\begin{center}
{\bf Figure captions}
\end{center}

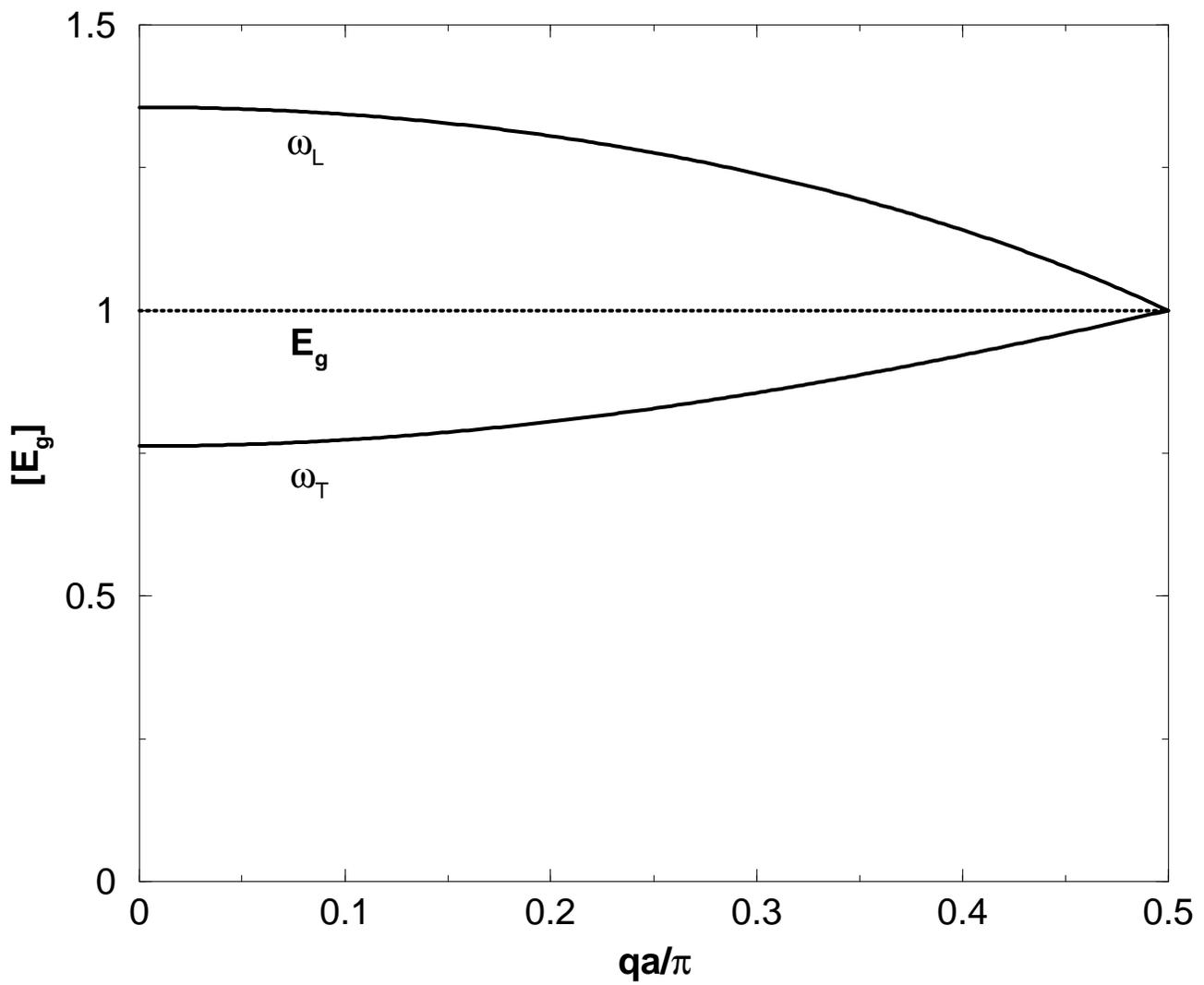
Fig.1. The band of collective dipolar  excitations for 
$16 \pi n_{e} \mu^{2}/(3a^{2}E_{g})=0.82$ and $U_{0011}=0$.

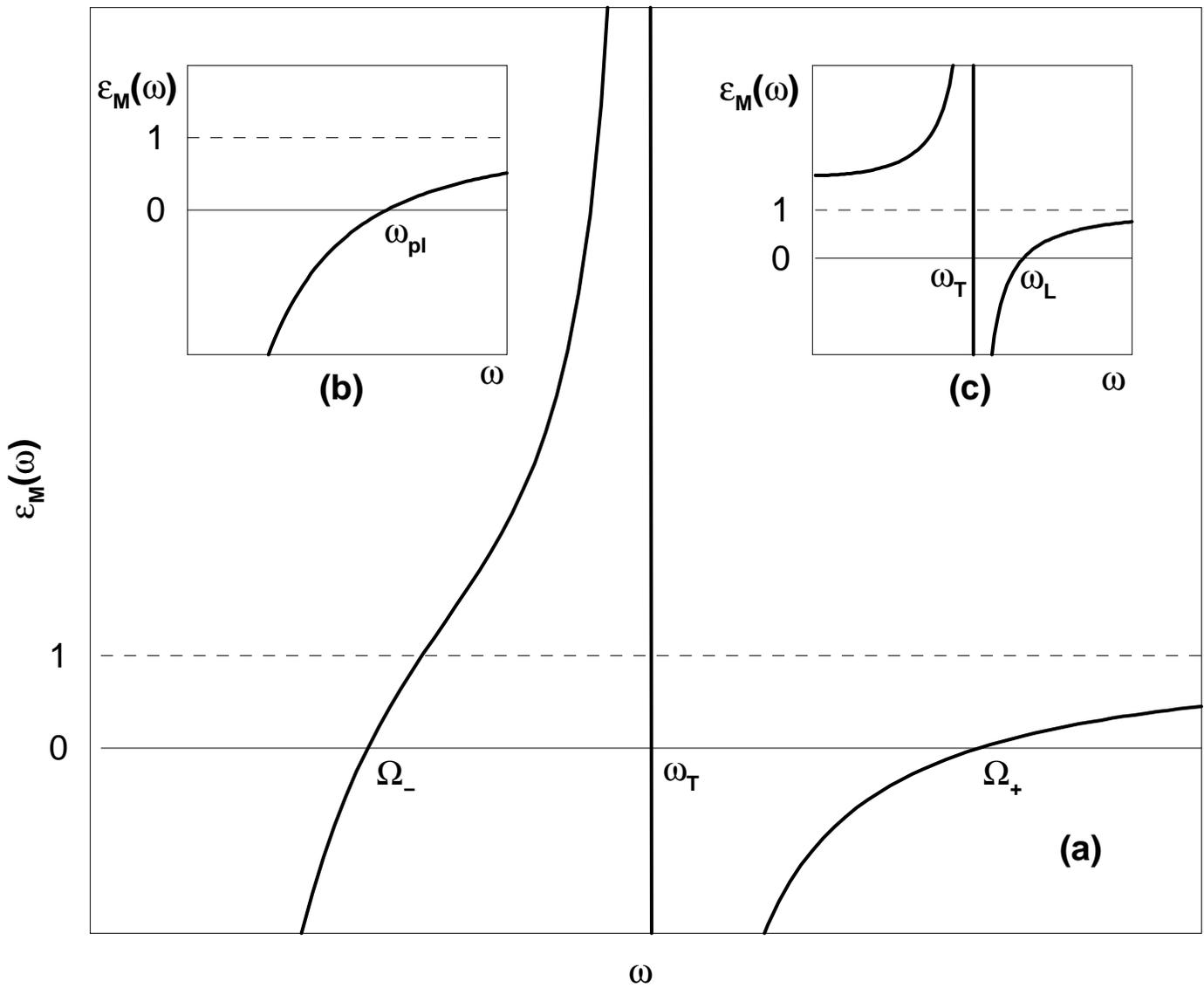
Fig.2. The schematic  frequency dependence of the macroscopic dielectric function 
 for  the (a) two band metal, (b) one band metal and (c) insulator.

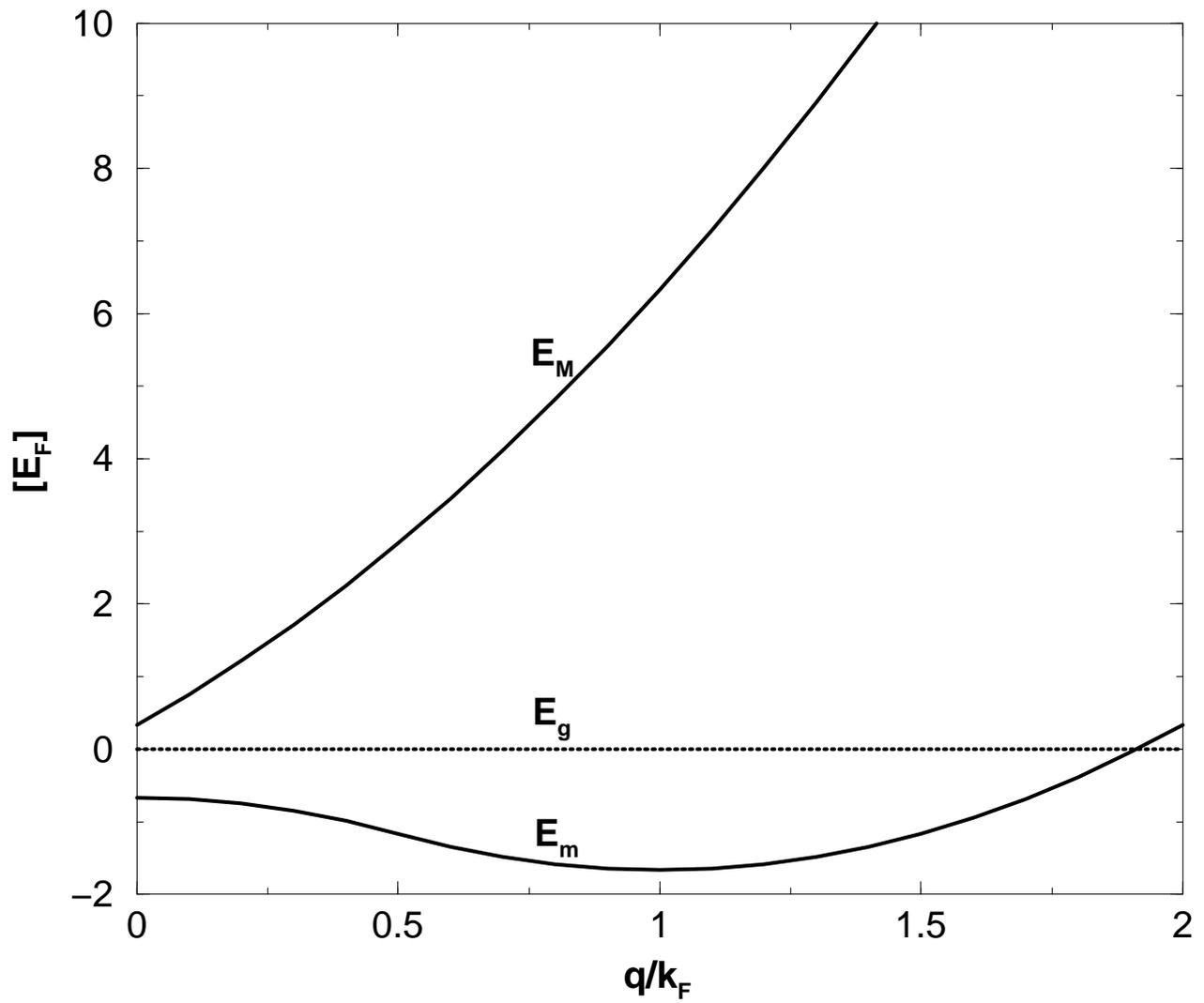
Fig.3. The region of interband  electron-hole  excitations for the band 
dispersion specified in front of eq.(\ref{aa2}). 

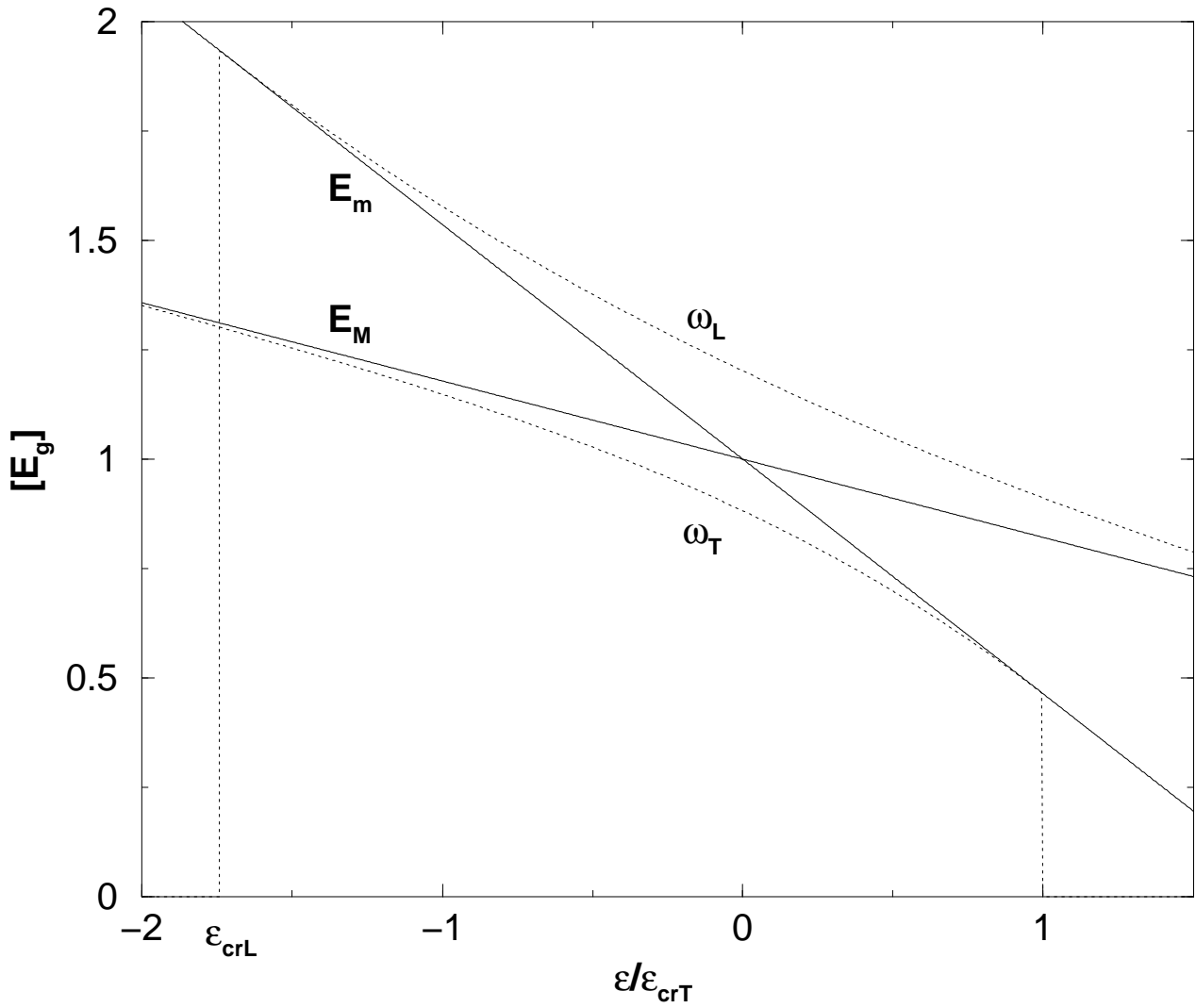
Fig.4. The frequencies of collective dipolar  excitations 
for {\bf q}=0 and the electron-hole continuum
{\em vs}  bandwidth difference  ${\cal E} \equiv (t_{1}-t_{0}) k_{F}^{2}a^{2}$
for a nearly empty valence band  and   $U_{0011}=0$.
Note   that  $| {\cal E}_{crL} | \neq 2 |{\cal E}_{crT} |$ since the
condition $E_{m} \gg {\cal E}$ is not fulfilled.

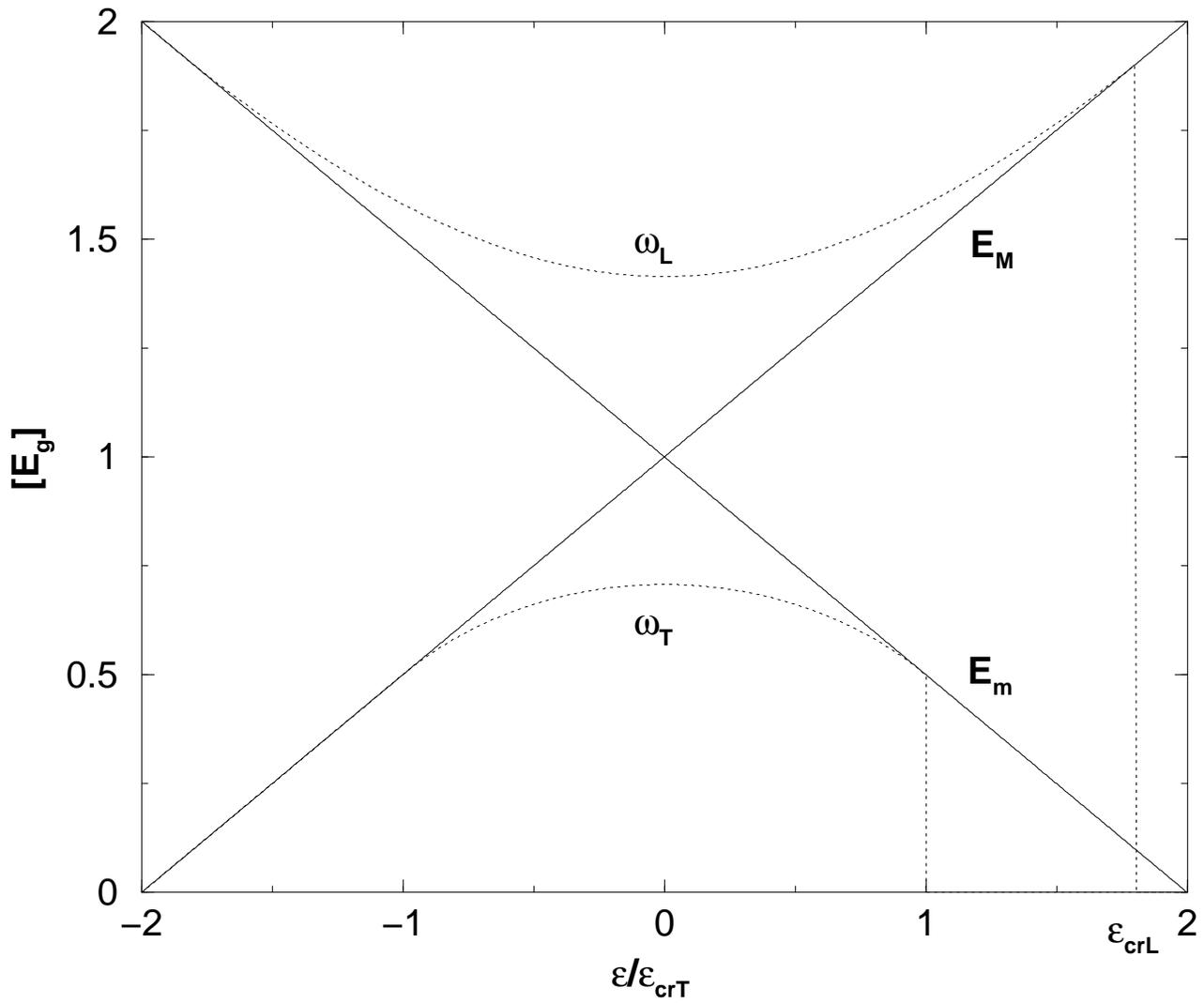
Fig.5. The frequencies  of dipolar excitations and the electron-hole continuum {\em vs}  
bandwidth difference ${\cal E}\equiv t_{1} - t_{0}$ for a  full valence band and 
$U_{0011}=0$. Note that  $| {\cal E}_{crL} | \neq 2 |{\cal E}_{crT} |$ since the
condition $E_{g} \gg {\cal E} $ is not fulfilled.

\end{sloppypar}
\end{document}